\documentclass[journal, 10pt, twocolumn]{IEEEtran}

\def\BibTeX{{\rm B\kern-.05em{\sc i\kern-.025em b}\kern-.08em
    T\kern-.1667em\lower.7ex\hbox{E}\kern-.125emX}}
\def\Var{{\rm Var}}
\def\Prob{{\rm Prob}}

\def\sgth{\sigma_\theta^2}

\def\Wsnr{{(1+\gamma_k^{-1})}}

\def\gikpp{(1+\gamma_k^{-1})}
\def\gik{1+\gamma_k^{-1}}
\def\gionepp{(1+\gamma_1^{-1})}

\newtheorem{lemma}{Lemma}[section]
\newtheorem{theorem}[lemma]{Theorem}

\newtheorem{remark}[lemma]{Remark}

\usepackage{times}
\usepackage{latexsym}
\usepackage{amsmath}
\usepackage{amsfonts}
\usepackage{pstricks}
\usepackage{amssymb}
\usepackage{amsxtra}
\usepackage{pstricks}
\usepackage{epsfig}
\usepackage{color}
\usepackage{graphics}
\usepackage{psfrag}
\usepackage{graphicx}
\usepackage{epstopdf}
\newcommand{\ie}{\emph{i.e.}}

\newcommand{\nr}{\nonumber}

\begin{document}

\title{Estimation Diversity and Energy Efficiency in Distributed Sensing}

\centerfigcaptionstrue


\author{Shuguang Cui, Jin-Jun Xiao,
Andrea Goldsmith, Zhi-Quan Luo, and H. Vincent Poor
\thanks{Part of this work was presented in
ICASSP'05 and ICC'06. This research was supported in part by funds
from the University of Arizona Foundation, the National Science
Foundation under grants No. CCR-02-05214 and No. DMS-0312416, the
U.S. Army MURI under award No. W911NF-05-1-0246, the Intel
Corporation, and the DOD ARMY under grant No. W911NF-05-1-0567.}
\thanks{S.\ Cui is with the Department of Electrical and Computer Engineering, University
of Arizona, Tucson, AZ 85721. Phone: 520-6269627. Fax: 520-6213862.
Email: cui@ece.arizona.edu.}
\thanks{A.\ J.\ Goldsmith is with the Wireless System Lab, Department of
Electrical Engineering, Stanford University, Stanford, CA 94305.
Phone: 650-7256932. Fax: 650-7235291. Email:
andrea@wsl.stanford.edu.}
\thanks{J.\ Xiao and Z.\ Q.\ Luo are with the Department of
Electrical and Computer Engineering, University of Minnesota,
Minneapolis, MN 55455. Phone: 612-6250242. Fax: 612-6254583. Emails:
{\textsf \{xiao,luozq\}@ece.umn.edu}.}
\thanks{H.\ V.\ Poor is with the Department of
Electrical Engineering, Princeton University, Princeton, NJ 08544.
Phone: 609-2581816. Fax: 609-2581468. Email: poor@princeton.edu.} }

\markboth{To appear at IEEE Transactions on Signal Processing, 2007}
{Cui \MakeLowercase{\textit{et al.}}: On the Estimation Diversity
and Energy Efficiency in Distributed Sensing}


%



\maketitle

\begin{abstract}

Distributed estimation based on measurements from multiple wireless
sensors is investigated.  It is assumed that a group of sensors
observe the same quantity in independent additive observation noises
with possibly different variances.  The observations are transmitted
using amplify-and-forward (analog) transmissions over non-ideal
fading wireless channels from the sensors to a fusion center, where
they are combined to generate an estimate of the observed quantity.
Assuming that the Best Linear Unbiased Estimator (BLUE) is used by
the fusion center, the equal-power transmission strategy is first
discussed, where the system performance is analyzed by introducing
the concept of estimation outage and estimation diversity, and it is
shown that there is an achievable diversity gain on the order of the
number of sensors. The optimal power allocation strategies are then
considered for two cases: minimum distortion under power
constraints; and minimum power under distortion constraints. In the
first case, it is shown that by turning off bad sensors, \ie,
sensors with bad channels and bad observation quality, adaptive
power gain can be achieved without sacrificing diversity gain. Here,
the adaptive power gain is similar to the array gain achieved in
Multiple-Input Single-Output (MISO) multi-antenna systems when
channel conditions are known to the transmitter. In the second case,
the sum power is minimized under zero-outage estimation distortion
constraint, and some related energy efficiency issues in sensor
networks are discussed.
\end{abstract}

\begin{keywords}
Estimation outage, estimation diversity, distributed estimation,
energy efficiency.
\end{keywords}

\section{Introduction}

\begin{figure}[!h]
\centering
      \scalebox{0.5}{\includegraphics{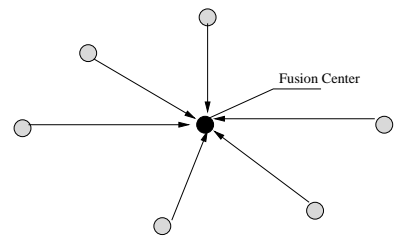}}
       \caption{Sensor network with a fusion center.}
    \label{wsn_graph}
\end{figure}

Wireless Sensor Networks (WSNs) deploy geographically-distributed
sensor nodes to collect information of interest. The collected
information is usually aggregated via wireless transmissions at a
fusion center to generate the final intelligence. A typical wireless
sensor network, as shown in Fig.~\ref{wsn_graph}, consists of a
fusion center and a number of sensors. The sensors typically have
limited energy resources and communication capability. Each sensor
in the network makes an observation of the quantity of interest,
generates a local signal (either analog or digital), and then sends
it to the fusion center where the received sensor signals are
combined to produce a final estimate of the observed quantity.
Sensor networks of this type are suited for various applications
such as environmental monitoring and smart factory instrumentation.

There has been a long history on the study of distributed
estimation. Examples of early work include the study in the context
of distributed control \cite{castanon, papadopoulos}, tracking
\cite{willsky}, or data fusion \cite{gubner,luo}. Recently, many new
results appear in the WSN community with a focus on distributed data
fusion, where the most commonly used network fusion model is the one
where each sensor processes its individual measurement and transmits
the result over a Multiple Access Channel (MAC) to the fusion
center. From an information-theoretic perspective,
\cite{gas_vet03,gas2,gas3,liu_sayeed,nowak2} investigate the mean
squared estimation error performance versus transmit power for the
quadratic CEO problem with a coherent MAC. Notably, it is shown in
\cite{gas_vet03} that if the sensor statistics are Gaussian, a
simple uncoded (analog-and-forwarding) scheme dramatically
outperforms the separate source-channel coding approach and leads to
an optimal asymptotic scaling behavior. The uncoded communication
scheme is further proved to preserve the optimal scaling law
in~\cite{liu_sayeed} for sensor networks with node statistics
satisfying a certain mean condition, while the source-channel
matching result is extended to more general homogeneous signal
fields in~\cite{nowak2}.  If the sensor measurements are not
continuous but in a finite alphabet, type-based transmission schemes
are proposed in~\cite{mergen_tong,liu_sayeed2}. The many-to-one
transport capacity and compressibility are investigated for dense
joint estimation sensor networks in~\cite{marco_neuhoff}. When a
full coordination among sensors is unavailable and the underlying
communication links are not reliable, the distributed estimation
problem is investigated in~\cite{ishwar}, where an
information-theoretic achievable rate-distortion region is elegantly
derived.
The work in \cite{nowak} studies the in-network processing
approaches based on a hierarchical data-handling and communication
architecture for the estimation of field sources. In addition, by
assuming only local sensor information exchange, \cite{xiao_boyd05}
proposes a distributed algorithm for reaching network-wide
consensus.



Among most of the existing studies, it is usually assumed that the
joint distribution of the sensor observations is known. However, in
some practical systems the probability density function (pdf) of the
observation noise is hard to characterize, especially for a large
scale sensor network. This motivates us to devise universal signal
processing algorithms that do not require the knowledge of the
observation noise distributions. Recently, universal Decentralized
Estimation Schemes (DESs) are proposed in~\cite{luo1}
and~\cite{luo3}. In~\cite{luo1}, the author considers the universal
distributed estimation in a homogeneous sensor network where sensors
have observations of the same quality, while in~\cite{luo3}, the
universal DES in an inhomogeneous sensing environment is considered.
These proposed DESs require each sensor to send to the fusion center
a short discrete message with length determined by the local Signal
to Noise Ratio (SNR), which then guarantees that the performance is
within a constant factor of that achieved by the Best Linear
Unbiased Estimator (BLUE). An assumption in these proposed schemes
is that the channels between the sensors and the fusion center are
perfect, \ie, all messages are received by the fusion center without
any distortion. However, due to power limitations, fading, and
channel noise, the signal sent by each individual sensor to the
fusion center may be corrupted. Therefore, the transmission system
for the joint estimation scheme should be designed to minimize the
end-to-end distortion subject to certain transmit power constraints,
under a practical wireless channel model considering both fading and
additive noises. In this paper we show that in a fading wireless
environment, multiple sensor nodes are not only necessary for
generating multiple observations to reduce distortion, but also
crucial to achieve a certain degree of diversity that minimizes the
effects of fading during signal transmissions.

If the sensor observation is in analog form, we have two main
options to transmit the observations from sensors to the fusion
center: analog or digital. For the analog approach, the observed
signal is transmitted via analog modulation to the fusion center,
which we refer to as the amplify-and-forward approach. In the
digital approach, the observed signal is digitized into bits,
possibly compressed and/or encoded, then digitally modulated and
transmitted. It is well known (\cite{Goblick},~\cite{Gastpar}) that
for a single Gaussian source with an Additive White Gaussian Noise
(AWGN) channel, both the digital and analog approaches can retain
the optimal power-distortion tradeoff. Also for the estimation of a
Gaussian source with a coherent Gaussian MAC, it is shown in
\cite{gas_vet03,liu_sayeed,nowak2} that the analog forwarding
schemes outperform (or are as good as) the digital approaches and
have the optimal asymptotic scaling behavior.
For sources with general distributions, type-based (each sensor
transmits the local type or histogram of its data in an analog
fashion over a MAC) parametric estimation schemes are proposed
in~\cite{liu_sayeed, mergen_tong,liu_sayeed2}.
In the above papers, the impact of coherent MAC schemes on the joint
source-channel optimality is discussed.

In this paper, instead of assuming a coherent MAC, we adopt
orthogonal channels between the sensors and the fusion center. The
main motivation for using orthogonal multiple access schemes such as
Frequency Division Multiple Access (FDMA) is the removal of the
requirement on the carrier-level synchronization among sensors (we
still require pair-wise synchronization between each sensor and the
fusion center). We assume that the observed signal is analog and the
observation noises are uncorrelated across sensors. In addition, we
assume that the second moments of the signal and noise are known to
the corresponding sensor and the fusion center. The fusion center
deploys the best linear unbiased estimator to generate estimates of
the unknown signal. In this setting, we investigate an analog
transmission system where observations are amplified and forwarded
to the fusion center. We first analyze the system performance in
fading channels by introducing the concept of estimation diversity.
We investigate the diversity gain that is achievable in a slow
fading environment, where it is assumed that the transmission
between sensors and the fusion center experiences i.i.d. fading
factors together with AWGNs. An outage is claimed if the end-to-end
distortion is larger than a certain threshold. In this case, we show
that using multiple sensors can achieve diversity to enhance the
outage performance, where the diversity order is equal to the total
number of sensors. We then find the optimal power allocation
strategy for the case where the end-to-end distortion is minimized
under certain transmit power constraints. The result leads to
turning off certain sensors with bad channels and bad observation
quality. By doing so, the achievable diversity order is not reduced
and extra adaptive power gain is obtained. We finally investigate
the converse problem to minimize the total power consumption under a
certain distortion constraint.

The rest of the paper is organized as follows. Section II discusses
the system model. Section III analyzes the distortion performance of
an equal-power transmission strategy, where the concept of
estimation diversity is introduced. Section IV addresses the case
where the transmission power is allocated in an optimal way to
minimize the distortion. Section V focuses on the converse case
where power is minimized subject to a distortion constraint. Section
VI summarizes the results and presents our conclusions.

\section{System Model}

We assume a sensor network with $K$ sensors where the observation
$x_k(t)$ at sensor $k$ is represented as a random signal $\theta(t)$
corrupted by observation noise $n_k(t)$: $x_k(t)=\theta(t)+n_k(t)$,
$t=0,1,2,\cdots$. We also assume that both $\theta(t)$ and $n_k(t)$
are i.i.d. over time $t$. Each sensor transmits the signal $x_k(t)$
to the fusion center where $\theta(t)$ is estimated from the
received version of $x_k(t)$'s, $k=1,\cdots,K$. We further assume
that $\theta(t)$ and $n_k(t)$ have zero mean and second moments
$\sigma_\theta^2$ and $\sigma_k^2$ respectively, based on which we
define the local observation SNR for sensor $k$ as:
$\gamma_k=\sgth/\sigma_k^2$.


We assume that $K$ sensors transmit their observations to the fusion
center via $K$ orthogonal channels (FDMA), where different channels
experience independent fading factors and zero-mean AWGNs.
Specifically, for channel $k$, we assume i.i.d. (over $t$) block
fading with the channel power gain denoted as $g_k(t)$, and i.i.d.
(over $t$) AWGNs denoted as $n_{ck}(t)$ of variance $\xi_k^2$,
$k=1,\cdots,K$, where the variances are assumed to be the same for
all $k$'s in this paper. We also assume pair-wise synchronization
between each sensor and the fusion center. However, synchronization
among sensor nodes is not required. At each sensor transmitter, we
adopt an analog amplify and forward uncoded strategy, motivated by
the results derived in~\cite{Gastpar}. Therefore, at sensor $k$, the
transmitter can be simply modeled by a power amplifying factor
$\alpha_k(t)$ and the average transmit power is given as
\begin{equation}\label{Eq_power1}
P_k=\alpha_k{P_{x_k}}=\alpha_k(\sgth+\sigma_k^2)
=\alpha_k'(1+\gamma_k^{-1})
\end{equation}
where $P_{x_k}$ is the average power of $x_k(t)$ and
$\alpha_k'=\alpha_k\sgth$. Note that we only need to consider the
power gains (no phase information is needed) in both the transmitter
and the channel, based on the assumptions that only the amplitude of
$\theta(t)$ is estimated and coherent reception (the effect of phase
is eliminated due to synchronization) is performed in the fusion
center for each $x_k(t)$.

\begin{figure}[!h]
\psfrag{theta}{$\theta$} \psfrag{b1}{$\sqrt{\alpha_1}$}
\psfrag{b2}{$\sqrt{\alpha_2}$}  \psfrag{bK}{$\sqrt{\alpha_K}$}
\psfrag{a1}{$\sqrt{g_1}$} \psfrag{a2}{$\sqrt{g_2}$}
\psfrag{aK}{$\sqrt{g_K}$} \psfrag{n1}{$n_1$} \psfrag{n2}{$n_2$}
\psfrag{n3}{$n_K$} \psfrag{nc1}{$n_{c1}$} \psfrag{nc2}{$n_{c2}$}
\psfrag{ncK}{$n_{cK}$} \psfrag{yk}{$y_1, \cdots, y_K$}
\begin{center}
\hspace{-45pt}
 {\includegraphics[width=3in, height=1.4in]{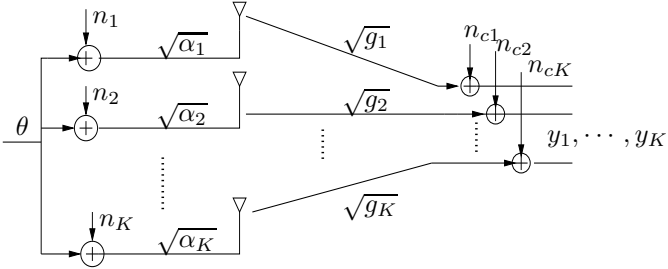}}
\end{center}
 \caption{Amplify and forward}\label{Fig_sensor_fa}

\end{figure}

Given the assumption of system independence over time $t$, we can
analyze the system performance by first focusing on an arbitrary
time snapshot, and then apply the result (which is conditional on
one system realization) to analyze the long-term average and outage
performance in the later sections. Therefore, from now on we neglect
the time index $t$ in all the parameters. The overall system
structure at one snapshot is shown in Fig.~\ref{Fig_sensor_fa}.

The received signal vector is given by
\begin{equation}
\mathbf{y}=\mathbf{h}\theta+\mathbf{v},
\end{equation}
where \vspace{-10pt}\begin{eqnarray} \mathbf{y}&=&\left[y_1,
y_2,\cdots,y_K\right]^{\dag}, \nonumber \\
\mathbf{h}&=&\left[{\sqrt{\alpha_1g_1}}, {\sqrt{\alpha_2g_1}},
\cdots, {\sqrt{\alpha_Kg_K}}\right]^{\dag}, \nonumber
\\ \mathbf{v}&=& \left[{\sqrt{\alpha_1g_1}}n_1+n_{c1},
\cdots,{\sqrt{\alpha_Kg_K}}n_k+n_{cK}\right]^{\dag}, \nonumber
\end{eqnarray}
with $\dag$ denoting transposition.

Since we intend to make the estimator universal (independent of
particular observation noise distributions except the second-order
statistics) and simple, the BLUE~\cite{Mendel1} is adopted at the
fusion center. Accordingly, the estimate for $\theta$ conditional on
a given set of channel gains is given by
\begin{eqnarray}
\hat{\theta}&=&[\mathbf{h}^{\dag}\mathbf{R}^{-1}\mathbf{h}]^{-1}\mathbf{h}^{\dag
}\mathbf{R}^{-1}\mathbf{y}
\nonumber \\
&=&\left(\sum_{k=1}^{K}\frac{{\alpha_kg_k}}{\sigma_k^2{\alpha_kg_k}+\xi_{k}^2}\right)^{-1}
\sum_{k=1}^{K}\frac{{\sqrt{\alpha_kg_k}}y_k}{\sigma_k^2{\alpha_kg_k}+\xi_{k}^2},
\end{eqnarray}
where the noise variance matrix $\mathbf{R}$ is a diagonal matrix
with $R_{kk}={\sigma_k^2{\alpha_k}{g_k}+\xi_{k}^2}$, $k=1,\cdots,K$.


The Mean Squared Error (MSE) of this estimator is given
as~\cite{Mendel1}
\begin{eqnarray}\label{Eq_distortion}
\Var[\hat{\theta}]
&=&[\mathbf{h}^{\dag}\mathbf{R}^{-1}\mathbf{h}]^{-1}\nr\\
&=&\left(\sum_{k=1}^{K}\frac{{\alpha_kg_k}}
{\sigma_k^2{\alpha_kg_k}+\xi_{k}^2}\right)^{-1}\nonumber \\
&=&\sgth\left(\sum_{k=1}^{K}\frac{{\alpha_k's_k}}
{\gamma_k^{-1}{\alpha_k's_k}+1}\right)^{-1},
\end{eqnarray}
where for notational convenience, we introduce the channel SNR:
$s_k={g_k}/{\xi_k^2}$, $k=1,\cdots,K$.

We now summarize the notations that we have defined so far (for a
particular time snapshot).
\begin{itemize}
\item $\theta$ and $\sigma_{\theta}^2$: Signal to estimate and its variance;
\item $n_k$ and $\sigma_k^2$: Observation noise and the
associated variance at sensor $k$;
\item $x_k$: Observation signal at sensor $k$;
\item $\alpha_k$: Power amplifying factor at sensor $k$;
In addition, $\alpha_k':=\alpha_k\sgth$;
\item $g_k$: Power gain of channel $k$;
\item $n_{ck}$ and $\xi_k^2$: Zero-mean AWGN of
channel $k$ and its variance, with $\xi_k^2$ the same for all $k$'s;
\item $\gamma_k:=\sgth/\sigma_k^2$: Local observation SNR at sensor $k$;
\item $s_k:={g_k}/{\xi_k^2}$: SNR of channel $k$.
\end{itemize}

\section{equal-power Allocation: Estimation Diversity}\label{Section_equalpower}

Given the proposed joint estimation system, we are interested in
investigating how the overall distortion performance is affected by
the fact that we have multiple sensors with independent fading
channels. We first investigate how the average distortion scales
with the number of sensors in the network, and secondly, we quantify
how the reliability of the overall estimation system is enhanced as
we increase the number of sensors given independent observations and
independent fading channels across different sensors.

To assure fairness when we compare different systems with different
numbers of sensor nodes, we fix the total transmission power that
the $K$ nodes can use, denoted as $P_{tot}$. In this section, we
consider the case of equal-power allocation where all sensors
transmit the same amount of power. As the total power budget for all
sensors is $P_{tot}$, according to Eq.~(\ref{Eq_power1}), we have
$$\alpha_k'=\frac{P_{tot}}{K\Wsnr}, \quad
1\le k\le K. $$
According to Eq.~(\ref{Eq_distortion}), the achieved MSE is
\begin{eqnarray}\label{mse_snr}
\Var[\hat{\theta}]
=\sgth\left(\sum_{k=1}^{K}\frac{{P_{tot}s_k}}
{\gamma_k^{-1}{P_{tot}s_k}+K(1+\gamma_k^{-1})}\right)^{-1}.
\end{eqnarray}

We assume that the channels between sensors and the fusion center
experience channel gain $g_k$'s, which are i.i.d. over $k$, and the
sensors have different observation noises with random variances
$\sigma_k^2$'s that are also i.i.d. over $k$. The i.i.d. assumption
on $\sigma_k^2$'s can be justified if we assume that the sensors are
randomly deployed into the field and the different measurement noise
variances are caused by different observation distances. With these
assumptions, we observe that both $\gamma_k$'s and $s_k$'s are
i.i.d. over $k$, and without loss of generality we have
$E(\gamma_k)=E(\gamma_1)$ and $E(s_k)=E(s_1)$, $k=1,\cdots,K$.

Our first question is: \textit{Suppose that the total power
$P_{tot}$ is fixed; what is the asymptotic behavior of
$\Var[\hat{\theta}]$ when the total number of sensors K increases
without bound}?

When $s_k$'s and $\gamma_k$'s are random and i.i.d. over $k$, we
have
\begin{eqnarray}\label{mse_aym_bounds_another}
\lim_{K \rightarrow \infty}\Var[\hat{\theta}]
=\frac{\sgth}{P_{tot}E[s_1/\gionepp]}:=D_\infty,
\end{eqnarray}
for which the derivation is given in Appendix A.

From the result in Eq.~(\ref{mse_aym_bounds_another}) and the
corresponding derivation given in Appendix A, we conclude the
following:
\begin{itemize}
\item With a finite amount of total transmit power $P_{tot}$ for all the sensors,
the overall MSE $\Var[\hat{\theta}]$ does not decrease to zero even
if $K$ approaches infinity. This is a consequence of using
orthogonal links from the sensors to the fusion center, which leads
to $K$ different channel noises such that the corruption of channel
noise cannot be eliminated even when $K$ goes to infinity. Systems
based on non-orthogonal multiple access schemes are discussed in
~\cite{analog_spawc,globecom06}, where it is shown that  with finite
total transmit power, $\Var[\hat{\theta}]$ can be driven to zero
when $K$ goes to infinity. However, in those systems perfect carrier
synchronization among all the sensor nodes and full channel
knowledge (amplitude gain and phase shift) at the transmitters are
required, which may not be feasible in practical systems.
\item Although $\Var[\hat{\theta}]$ is bounded away from zero, it
decreases monotonically with $K$. However, the reduction in
distortion with each additional sensor decreases as $K$ gets large
(c.f. Eq.~(\ref{mse_aym_bounds}));
\item When $K$ is large, $\Var[\hat{\theta}]$ is inversely proportional to
$P_{tot}$. Thus, when $K$ increases, if $P_{tot}$ also increases (at
any speed) with $K$, we have $D_{\infty}=0$.
\end{itemize}

This analysis suggests that when the total amount of power $P_{tot}$
is fixed, even though the total number of sensors $K$ increases
without bound, the achieved average distortion at the fusion center
does not decrease below a certain level $D_{\infty}$. However, are
there any benefits of having more sensors in the network if we limit
the amount of total power? To answer this question, let us define
the outage probability ${\mathcal{P}_{D_0}}$ to model the system
reliability as follows,
\begin{equation}\label{outage_prob}
\mathcal{P}_{D_0}=\Prob\{\Var[\hat{\theta}]>D_0\},
\end{equation}
where $D_0$ is a predefined threshold. Given the i.i.d. nature of
$s_k$'s and $\gamma_k$'s, the probability of
$\Var[\hat{\theta}]>D_0$ at one particular snapshot is an
appropriate indicator of the long-term estimation system
reliability. The following theorem summarizes the relationship among
$\mathcal{P}_{D_0}$, $P_{tot}$, and the number of sensors $K$.

\vspace{2mm}

\begin{theorem}\label{main_theorem}

Suppose sensor observation SNR $\{\gamma_k:k=1,2,\ldots,K\}$ and
channel SNR $\{s_k:k=1,2,\ldots,K\}$ are both i.i.d. across $k$.
Define $\eta_k:=s_k/\gikpp$. In addition, we assume that $E[\eta_k]$
and $E[\gamma_k^{-1}{s_k^2}]$ are finite. When the total number of
sensors $K$ is large, with the total power $P_{tot}$ and equal-power
allocation among sensors, we have the achievable average distortion
$$D_\infty:=\lim_{K\rightarrow \infty}\Var[\hat{\theta}]
=\frac{\sgth}{P_{tot}E[\eta]}.$$ Moreover, for a sufficiently large but finite $K$ and
$D_0>D_\infty$\footnote{For the other cases of $D_0\le{D_\infty}$,
it is easy to see that ${\mathcal{P}_{D_0}}\to 1$.}, we have the
outage probability (c.f. Eq.~(\ref{outage_prob}))
\begin{eqnarray}
{\mathcal{P}_{D_0}}\sim\exp(-KI_\eta(a)),\hspace{0.2cm}\mbox{or}
\hspace{0.2cm} -\log{\mathcal{P}_{D_0}}\sim{KI_\eta(a)}, \nr
\end{eqnarray}
where $\sim$ means asymptotically converging to (as $K$ becomes
large), $\eta$ is the common distribution of $\eta_k$,
$a=\sgth/(D_0P_{tot})$, and $I_\eta(a)$ is the rate function of
$\eta$:
\begin{eqnarray}
I_\eta(a)=\sup_{\theta\in\mathbb{R}}(\theta a-\log
M_\eta(\theta)),\nr
\end{eqnarray}
with $M_\eta(\theta)$ the moment generating function of $\eta$.

\end{theorem}

\vspace{2mm}

A more detailed explanation of the rate function and the proof of
Theorem~\ref{main_theorem} are given in Appendix B.

From the theorem we see that $K$ plays the role of estimation
diversity order here, in that the outage probability decreases
exponentially with K. We remark that the fact that the outage
probability decays exponentially with the number of sensors is due
to the effect of independent measurements and fading coefficients,
which bears similar properties as the probability of detection error
in distributed detection \cite{sayeed, chamberland, liu_sayeed2,
xiao_dds}. Note that even though Theorem~\ref{main_theorem} is an
asymptotic result for large $K$, we later show by simulation results
that the outage probability curve illustrates diversity order of $K$
(approximately) even for small values of $K$ in practical scenarios.

As an example, let us consider the case in which $\gamma_k=1$ for
$k$'s and $\sqrt{s_k}$ is i.i.d. Rayleigh with pdf
\begin{eqnarray}
f_{\sqrt{s}}(x)=\frac{x}{\delta^2}\exp\left\{-\frac{x^2}{2\delta^2}\right\}.\nr
\end{eqnarray}
Then $\eta_k=s_k/2$ has exponential distribution with pdf
\begin{eqnarray}
f_\eta(x)=\frac{1}{\delta^2}\exp\left\{-\frac{x}{\delta^2}\right\}.\nr
\end{eqnarray}
Therefore, $E[\eta_k]=\delta^2$. Thus the achieved asymptotic
distortion when $K$ is large is given by
\begin{eqnarray}
D_{\infty}=\lim_{K\rightarrow
\infty}\Var[\hat{\theta}]=\frac{\sgth}{P_{tot}E[\eta_k]}=
\frac{\sgth}{P_{tot}\delta^2}.\nr
\end{eqnarray}

Now we calculate the rate $I_{\eta}(a)$. It is easy to see that the
moment generating function of an exponentially distributed random
variable with mean $b$ is given as
\begin{eqnarray}
M(\theta)=\frac{1}{1-b\theta}.\nr
\end{eqnarray}
Thus
\begin{eqnarray}
I_{\eta}(a)&=&\sup_{\theta\in\mathbb{R}}\left[\theta
a+\log(1-b\theta)
\right]\nr\\
&=&\frac{a}{b}-\log\frac{a}{b}-1\nr\\
&=&\frac{\sgth}{\delta^2D_0P_{tot}}-\log
\frac{\sgth}{\delta^2D_0P_{tot}}-1,\nr
\end{eqnarray}
where in the last step, we substituted the expressions
$a=\sgth/(D_0P_{tot})$ and $b=\delta^2$. When $P_{tot}$ is large,
\ie, when $\displaystyle\frac{\sgth}{\delta^2D_0P_{tot}}\ll 1$, this
leads to
\begin{eqnarray}
I_{\eta}(a)&=&\frac{\sgth}{\delta^2D_0P_{tot}}-\log
\frac{\sgth}{\delta^2D_0P_{tot}}-1\nr\\
&\sim&\log P_{tot} ,\nr
\end{eqnarray}
which means that the estimation convergence rate is approximately
$\log{P_{tot}}$ when
$\displaystyle\frac{\sgth}{\delta^2D_0P_{tot}}\ll 1$ is satisfied.
In other words, for Rayleigh fading channels we have
\begin{equation}\label{Eq_diversity_Rayleigh}
-\log {\mathcal{P}_{D_0}} \sim K \log P_{tot},
\end{equation}
which shows that the diversity order $K$ is the slope of the outage
probability vs. power curve if things are plotted in the log-log
fashion.

%

\vspace{2mm}

We now provide some numerical examples to verify the analytical
results. We assume that the channel SNR is given as
$s_k=\frac{G_0}{\xi_k^2d_k^{2}}|r_k|^2$ where $d_k$ is the
transmission distance from sensor $k$ to the fusion center,
$G_0=-30$~dB is the nominal gain at $d_k=1$~m, and the $|r_k|$'s are
i.i.d. Rayleigh fading random variables with unit variance. We take
$\xi_k^2=-90$~dBm, $k=1,\cdots,K$. To emphasize the possible
diversity gain enabled by the independent channel fading values, we
set $d_k=100$~m and $\sigma^2_k=0.01$ for all $k$. In addition, we
set $\sigma_{\theta}^2=1$. The outage threshold $D_0$ is set as
$D_0=2\sigma^2_k=0.02$.

\begin{figure}[!h]
\centering
      \scalebox{0.55}{\includegraphics{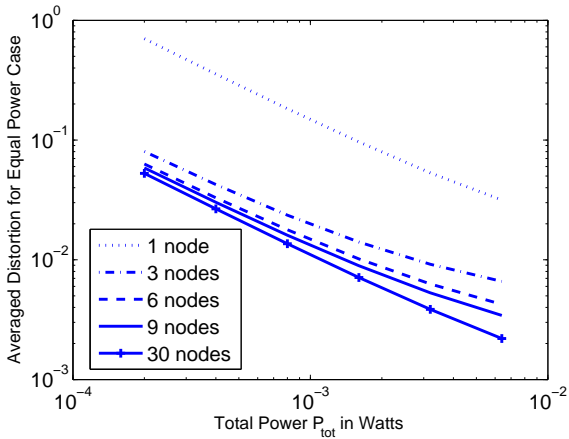}}
       \caption{Average Distortion vs. Total Power}
    \label{distortion}
\end{figure}

The end-to-end distortion performance, averaged over random channel
gains, is plotted in Fig.~\ref{distortion} for different numbers of
sensors, where each point is a sample average over one million
independent random channel samples. It is not surprising that the
average distortion decreases as we increase the total power budget.
Note that the average distortion barely improves when we increase
the number of nodes from $3$ to $30$, which matches the comments
given after Eq.~(\ref{mse_aym_bounds_another}). However, this does
not mean that the 3-node case performs as well as the 30-node case,
since the average performance is not a good criterion to use in a
slow fading environment, where the outage performance is more
informative.

\begin{figure}[!h]
\centering
      \scalebox{0.55}{\includegraphics{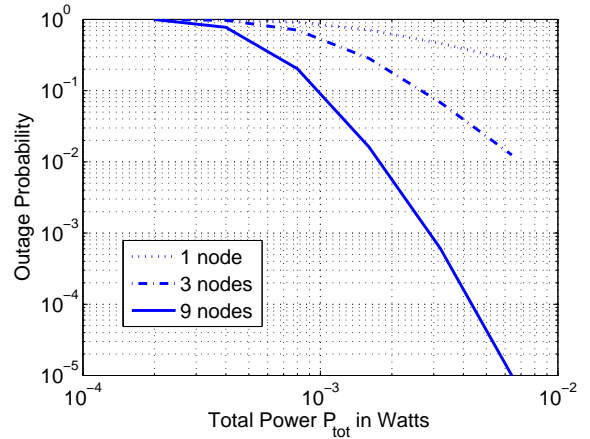}}
       \caption{Outage Probability vs. Total Power}
    \label{outage}
\end{figure}

\begin{figure}[!h]
\centering
      \scalebox{0.55}{\includegraphics{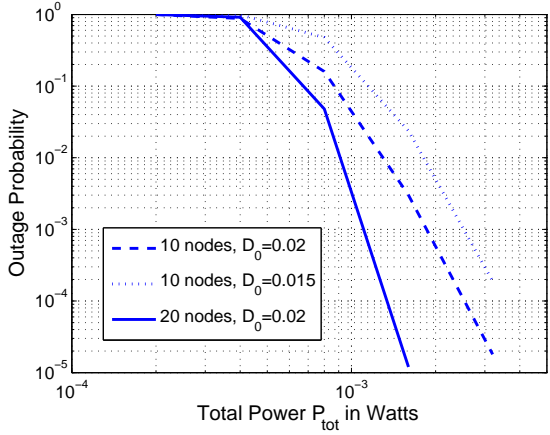}}
       \caption{Outage Probability vs. Total Power for Large Numbers of Sensors}
    \label{outage_largeN}
\end{figure}

The outage probability versus the total transmission power is
plotted in Fig.~\ref{outage} for different numbers of sensors, where
we see that the 3-node case performs much better than the 1-node
case and the 9-node case performs much better than the 3-node case.
Approximately, when the logarithm of outage probability is plotted
versus the logarithm of total transmission power, the slope of the
curve at the high power region is proportional to the number of
sensors, which is defined as the diversity order in
Eq.~(\ref{Eq_diversity_Rayleigh}). Note that this definition of
diversity order is based on the distortion outage performance, which
is different from the traditional definition of diversity order in
Multiple-Input Multiple-Output (MIMO) multi-antenna
systems~\cite{Paulraj1}, which is usually the slope of symbol error
curves. However, the two definitions imply similar performance
benefits from diversity. This type of diversity gain is also shown
for large numbers of sensors in Fig.~\ref{outage_largeN}, where we
see that for the same $D_0=0.02$ threshold, the slope of the 20-node
curve is twice that of the 10-node curve in the high power region.
Not surprisingly, when we decrease $D_0$ (down to $0.015$ as shown
in Fig.~\ref{outage_largeN}), the outage probability will be
increased. It is worth mentioning that since $D_{\infty}$ decreases
with $P_{tot}$, when $P_{tot}$ increases, a fixed $D_{0}$ becomes
progressively conservative as it gets further away from
$D_{\infty}$. As such, a more appropriate definition for the outage
probability may be $P_{\epsilon}=\Prob\{\Var[\hat{\theta}]
>(1+\epsilon)  D_{\infty}\}$ (as pointed out by one of the reviewers),
which is definitely worth further investigation, but beyond the
scope of this paper.

\section{Optimal power Allocation: Diversity Gain + Power Gain}\label{Section_min_distortion}

In the previous section we showed that diversity gain can be
achieved even if we use a simple uniform power allocation scheme. In
this section, we optimize power allocation among the sensors to
minimize the total distortion. The diversity performance is analyzed
and we show that a certain adaptive power gain can be achieved by
optimal power control. 
To clarify the analysis, we first
discuss the problem with only a sum power constraint, then discuss
the general case with both sum and individual power constraints.

\subsection{Optimal power allocation with a sum power constraint}

With a sum power constraint, the minimum distortion joint estimation
problem for each given set of channel gains can be cast as
\begin{eqnarray}
\min && \Var[\hat{\theta}] \nr
\\ \mbox{s.~t.} &&
\sum_{k=1}^{K}P_k \le P_{tot}, \nr
\end{eqnarray}
where $P_{tot}$ is the total power constraint across all the nodes.
With Eqs.~(\ref{Eq_power1}) and (\ref{Eq_distortion}), we can
rewrite the above problem as
\begin{eqnarray} \min_{\alpha_k'} && \sgth\left(\sum_{k=1}^{K}
\frac{{\alpha_k's_k}}{\gamma^{-1}\alpha_k's_k+1}\right)^{-1}\nr
\\ \mbox{s.~t.} &&
\sum_{k=1}^{K}\alpha_k' \Wsnr \le{P_{tot}} \nonumber
 \\ && {\alpha_{k}'}\ge0,
\hspace{0.3cm}k=1,\cdots,K.\nr
\end{eqnarray}
Our goal is to obtain the optimal power allocation, i.e., optimal
$\alpha'_k$'s. To simplify the objective function, we rewrite the
problem as
\begin{eqnarray}\label{Eq_opt}
\min_{\alpha_k'} &&
-\sum_{k=1}^{K}\frac{{\alpha_k's_k}}{\gamma_k^{-1}{\alpha_k's_k}+1}
\nr
\\ \mbox{s.~t.} &&
\sum_{k=1}^{K}\alpha_k'\Wsnr \le{P_{tot}},
 \\ && {\alpha'_{k}}\ge0,
\hspace{0.3cm}k=1,\cdots,K.\nr
\end{eqnarray}
For nonnegative $\alpha_k'$, it can be shown that the second
derivative with respect to $\alpha_k'$ of each item in the objective
function is nonnegative. Since the objective function is also
separable over $\alpha_k'$ (no coupled terms over different $k$'s),
it is jointly convex over all the $\alpha'_k$'s. In addition, all
the constraints are linear constraints. Thus, the problem is convex.

Now we solve the problem in Eq.~(\ref{Eq_opt}). Its Lagrangian $G$
is given as
\begin{eqnarray}
G(\alpha_k';\lambda_0,\mu_k)&=&-\sum_{k=1}^{K}
\frac{{\alpha_k's_k}}{\gamma_k^{-1}{\alpha_k's_k}+1} \nr\\&&
-\lambda_0\left({P_{tot}}- \sum_{k=1}^K\alpha_k'\Wsnr\right)
-\sum_{k=1}^{K}\mu_k\alpha'_k\nr
\end{eqnarray}
which leads to the following Karush-Kuhn-Tucker (KKT)
conditions~\cite{Boyd1}:
\begin{eqnarray}
-\frac{s_k^{-1}}{\left(\gamma_k^{-1}\alpha'_k+s_k^{-1}\right)^2}
+\lambda_0\Wsnr-\mu_k=0, && \forall{k},\nr\\
\sum_{k=1}^K\alpha_k'\Wsnr-{P_{tot}}=0,\nr
&&\\
\mu_k\alpha_k'=0 && \forall{k},\nr \\
\mu_k\ge0 && \forall{k},\nr \\
\alpha_k'\ge0 && \forall{k}.\nr
\end{eqnarray}

From the first equation in the above set we obtain
\begin{eqnarray}
-\frac{s_k^{-1}}{\left(\gamma_k^{-1}\alpha'_k+s_k^{-1}\right)^2}
+\lambda_0\Wsnr=\mu_k,\nr
\end{eqnarray}
which leads to the solution
\begin{eqnarray}
\alpha_k' &=&\frac{\gamma_k}{s_k}
\left(\frac{1}{\sqrt{s_k^{-1}\left(\lambda_0\Wsnr
-\mu_k\right)}}-1\right).\nr
\end{eqnarray}
Also we can see from the third equation that for those sensors with
$\alpha_k'>0$ (i.e., $P_k>0$), $\mu_k=0$ holds. Therefore,
\begin{eqnarray}\label{alphak}
\alpha_k' &=&\frac{\gamma_k}{s_k}
\left(\frac{1}{\sqrt{s_k^{-1}\lambda_0\Wsnr}}-1\right)^+\nr\\
&=&\frac{\gamma_k}{s_k}
\left(\sqrt{\frac{\eta_k}{\lambda_0}}-1\right)^+, \forall{k},
\end{eqnarray}
where $(x)^+$ equals $0$ when $x$ is less than zero, and otherwise
equals $x$. The first equality follows from the fact that if
$\alpha_k'>0$, $\mu_k=0$, and if $\alpha_k'=0$, then removing
$\mu_k$ results in the difference within the bracket being
non-positive.

The Lagrangian multiplier $\lambda_0$ in Eq.~(\ref{alphak}) and the
number of active sensors (that are assigned non-zero power) can be uniquely
determined from the power constraint by the following two-step derivation.

\begin{itemize}

\item First, let us assume that only the first $K_{1}$ sensors are
active such that $\lambda_{0}$ can be solved by substituting
$[\alpha_{1}^{\prime},\cdots, \alpha_{K_{1}}^{\prime}]$ back to the
second KKT condition. This assumption can be guaranteed by ranking
the sensors (according to $\eta_{k}$ that is a function of both the
observation SNR and channel SNR) such that
\begin{eqnarray}\label{rank}
\eta_1\ge\eta_2\ge\ldots\ge\eta_K,
\end{eqnarray}
and the fact that $\alpha_{k}^{\prime}=0$ if $\eta_{k}\le \lambda_{0}$. As such, we obtain
$$\lambda_0=
\left(\frac{A(K_1)}{B(K_1)}\right)^2,$$ where for any $1\le k\le K$,
\begin{eqnarray}
A(k)&=&\displaystyle\sum_{m=1}^{k} \frac{\gamma_m}{\sqrt{\eta_m}}
\nr\\
B(k)&=&\displaystyle\sum_{m=1}^{k} \frac{\gamma_m}{\eta_m}+P_{tot}.
\nr
\end{eqnarray}

\item Secondly, we substitute $\lambda_{0}$ back to Eq.
(\ref{alphak}) and solve the cutoff index $K_1$, which is obviously
determined by the relative magnitudes between
$\sqrt{\frac{\eta_{k}}{\lambda_{0}}}$ and $1$. Naturally, we
introduce the notation:
\begin{eqnarray}\label{Eq_threshold1}
f(k)&=&\sqrt{\frac{\eta_{k}}{{\lambda_{0}}}}-1 \nonumber \\
&=&\frac {\sqrt{\eta_{k}}B(k)}{A(k)}-1,\quad \mbox{for\
}1\le{k}\le{K}.
\end{eqnarray}
%
%
It follows from Eqns. (\ref{alphak}) and (\ref{Eq_threshold1}) that
solving the threshold $K_1$ is equivalent to finding $K_1$ such that
$f(K_1)>0$ and $f(K_1+1)\le0$. Using the same techniques as
in~\cite[Appendix B]{Jinjun_Shuguang1}, we can show that such a
$K_1$ is unique and always exists unless $f(k)>0$ for all $1\le k\le
K$, in which case we take $K_1=K$ that means all sensors being
active.

\end{itemize}

Hence, it follows from Eq.~(\ref{alphak}) that
\begin{eqnarray}\label{Eq_powerallocation1}
{\alpha'_k}^{opt}=\left\{\begin{array}{ll}0, & k>K_1\\
\displaystyle\frac{\gamma_k}{s_k} \left(\sqrt{\eta_k}c_0-1\right),
&k\le K_1,\end{array}\right.
\end{eqnarray}
where $c_0=\sqrt{\lambda_0^{-1}}=B(K_1)/A(K_1)$. It is easy to see
that $c_0$ defines the threshold of the $\eta_k$'s (\ie,
$\eta_k\ge{1/c_0^2}$), by which we can decide whether a sensor
transmits or keeps silent. Note that the figure of merit
$\eta_k=s_k/(1+\gamma_k^{-1})$ is jointly defined by the channel SNR
and local observation SNR. For sensors with low $\eta_k$, they are
completely shut off and no power is wasted. For the remaining active
sensors, power should be assigned according to
Eq.~(\ref{Eq_powerallocation1}).


To implement the described optimal power scheduling scheme, the
fusion center needs to calculate and broadcast the threshold
$\lambda_0$ to all the sensors. Each sensor then decides the optimal
transmit power according to its own local information ($\gamma_k$
and $s_k$) and $\lambda_0$. Such a power scheduling scheme is
feasible when there exists a feedback broadcast channel (of low
rate) from the fusion center to each sensor and the channel changes
slowly.

Furthermore, according to Eq.~(\ref{Eq_distortion}), the total
distortion is given by
\begin{equation}\label{eq_dis}
\Var[\hat{\theta}]=\sgth\left(\sum_{k=1}^{K_1}\gamma_k\left(1-\frac{1}{c_0\sqrt{\eta_k}}
\right)\right)^{-1},
\end{equation}
and the outage probability can be rewritten as
\begin{equation}
{\mathcal{P}_{D_0}}=\Prob\left\{\sum_{k=1}^{K_1}\gamma_k\left(1-\frac{1}{c_0\sqrt{\eta_k}}
\right) <\frac{\sgth}{D_0}\right\}.
\end{equation}

To obtain closed-form representations for the outage probability is
difficult. However, we can numerically evaluate the performance for
the optimal power transmission schemes, and compare it with the
closed-form solution developed for the equal-power case in the
previous section. Since equal-power allocation is just one feasible
solution of the optimization problem in Eq.~(\ref{Eq_opt}), the
resulting optimal solution (which may turn off bad sensors) leads to
strictly-lower distortion than the equal-power allocation strategy.
Given that we have theoretically shown that the equal-power
allocation strategy achieves full estimation diversity (on the order
of $K$), we can state that the optimal power allocation strategy
performs at least equally well, {\it i.e.}, achieves full estimation
diversity. This is further illustrated by the following simulation
results.

\begin{figure}[!h]
\centering
      \scalebox{0.55}{\includegraphics{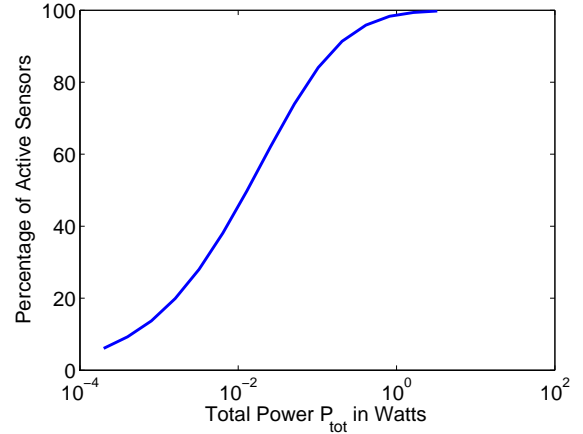}}
       \caption{Percentage of Active Sensors vs. Total Power}
    \label{active_sensor}
\end{figure}

\begin{figure}[!h]
\centering
      \scalebox{0.55}{\includegraphics{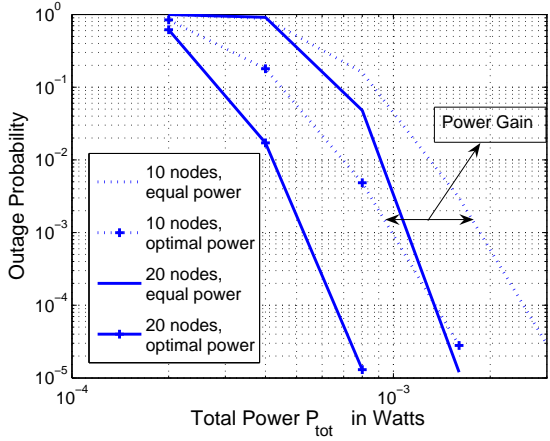}}
       \caption{Outage Probability Comparison vs. Total Power}
    \label{outage_compare}
\end{figure}

We assume that the related system parameters are set the same as in
the equal-power case in Section~\ref{Section_equalpower}. In
Fig.~\ref{active_sensor}, we plot the percentage of active sensors
versus the total transmission power, where we set $K=100$ in the
simulation. We note that the number of active sensors can be less
than $K$ when the total power budget is small. In
Fig.~\ref{outage_compare}, we compare the outage performance of the
optimal power scheme with the case where all the sensors transmit
with equal power. From the outage probability curves, we see that
for the same number of sensor nodes the curve slopes are almost the
same for both the equal-power and the optimal power cases, which
means that the optimal power transmission strategy achieves the same
diversity order of $K$. In addition, the curve for the optimal power
case is a left-shifted version of that for the equal-power case.
This shift is a result of adaptive power gain that is due to the
optimal power control.
This gain is similar to array or coding gain in traditional MIMO
systems~\cite{Paulraj1}.

\subsection{Optimal power allocation with both sum and individual power
constraints}

In the optimization problem of Eq.~(\ref{Eq_opt}), the sum power
constraint is imposed to guarantee a fair comparison when we change
the number of sensor nodes. In some application scenarios, this sum
power constraint has a physical meaning. For example, let us assume
that there are multiple clusters of sensors, where each cluster is
performing a different observation task. If different clusters are
sharing the same frequency band to transmit information, the total
power emitted from each cluster must be limited to enable the
coexistence of multiple clusters. In addition, a more severe power
constraint may be imposed on each individual sub-band used by each
sensor for better frequency reuse, which is modeled by individual
power constraints for all the sensors. Note that the individual
power constraint may also be imposed by power supply characteristics
at each node.

Given these considerations, we now cast an optimization model with
both individual and sum power constraints. The optimization problem
then becomes
\begin{eqnarray}
\min_{\alpha_k'} &&
-\sum_{k=1}^{K}\frac{{\alpha_k's_k}}{\gamma_k^{-1}{\alpha_k's_k}+1}
\nr
\\ \mbox{s.~t.} &&
\sum_{k=1}^{K}\Wsnr\alpha_k' \le{{P_{tot}}},
\\ && \alpha_k'\Wsnr \le {P_k^{max}}, \hspace{0.3cm}k=1,\cdots,K \nr
 \\ && {\alpha'_k}\ge0,
\hspace{0.3cm}k=1,\cdots,K \nr
\end{eqnarray}
where $P_k^{max}$ is the maximum allowable transmit power for node
$k$. By combining the last two sets of constraints, the problem can
be simplified to
\begin{eqnarray}\label{Eq_opt_complete}
\min_{\alpha_k'} &&
-\sum_{k=1}^{K}\frac{{\alpha_k's_k}}{\gamma^{-1}{\alpha_k's_k}+1}
\nr
\\ \mbox{s.~t.} &&
\sum_{k=1}^{K}\alpha_k'\Wsnr \le{{P_{tot}}},
\\ && 0\le {\alpha_k'} \le {C_k}, \hspace{0.3cm}k=1,\cdots,K \nr
\end{eqnarray}
where $C_k:={P_k^{max}}/{\Wsnr}$.

The optimization problem given in Eq.~(\ref{Eq_opt_complete}) is
still convex, since we have only added extra linear constraints into
the problem of Eq.~(\ref{Eq_opt}). However, it is now more difficult
to compute an analytical solution. We propose the following
algorithm to derive the optimal analytical solution.

\vspace{10pt} \noindent {\bf The Algorithm:}
\begin{enumerate}
\item Solve the problem without individual power constraints (\ie,
Eq.~(\ref{Eq_opt})) to obtain the solution as in
Eq.~(\ref{Eq_powerallocation1}); \\
Set the index set $\mathcal{K}_e=\{k|{\alpha'_k}^{opt}\ge{C_k}\}$.

\item Set ${\alpha'_k}^{opt}=C_k$ for $k\in{\mathcal{K}_e}$; \\
Set $P_{tot}=P_{tot}-\sum_{k\in{\mathcal{K}_e}}\Wsnr C_k$; \\
Remove $\alpha'_k$ for $k\in{\mathcal{K}_e}$ from the design
variable space.

\item Repeat the previous two steps until $\mathcal{K}_e$ is empty in Step (1).
\end{enumerate}

To prove that the proposed algorithm leads to the global optimum, we
only need to prove that in Step (2) we do not lose optimality of
${\alpha'_k}^{opt}$ for $k\in{\mathcal{K}_e}$ when we set
${\alpha'_k}^{opt}=C_k$ for $k\in{\mathcal{K}_e}$. This can be shown
easily by noticing that the objective function is monotonically
decreasing with respect to $\alpha_k'$ for all $k$. Since in Step
(2) we assign the maximum allowable values to ${\alpha'_k}^{opt}$
for $k\in{\mathcal{K}_e}$ within the feasible region, there is no
optimality loss, \ie, they are assigned the optimal values that
minimize the objective function.

\begin{figure}[!h]
\centering
      \scalebox{0.55}{\includegraphics{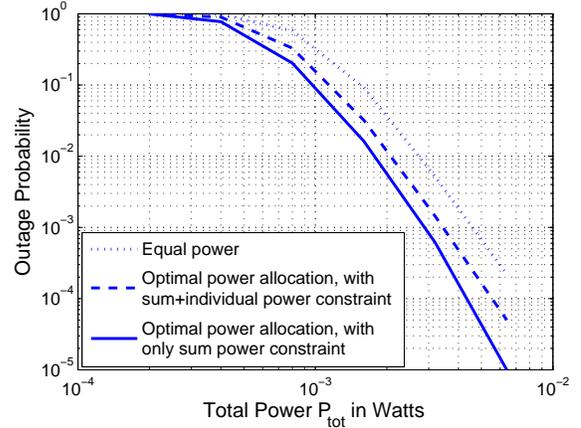}}
       \caption{Effect of Individual Power Constraints (6 nodes)}
    \label{outage_compare_Pk_6node}
\end{figure}

To illustrate how the individual power constraints affect the outage
performance, we take a six-node example. The other parameters are
set the same as before. We plot the outage performance in
Fig.~\ref{outage_compare_Pk_6node}, where each node has an
individual power constraint $P_k^{max}={(1.5P_{tot})}/{K}$ in
addition to the sum power constraint. From the curves we see that
the diversity order is kept the same when we have individual power
constraints, but the adaptive power gain over the equal-power case
is reduced compared with the case where we only have a sum power
constraint.

\subsection{Practical Issues}
To obtain the desired transmit power levels for each sensor, we have
assumed that the fusion center knows
$\{(\gamma_k,s_k):k=1,2,\ldots,K\}$. This assumption is reasonable
in cases where the network condition and the signal being estimated
change slowly in a quasi-static manner. We have also assumed that
the fusion center executes the optimization and then appropriately
activates the sensors with their respective power levels. Our
approach is general for the estimation of a memoryless discrete-time
random process $\theta(t)$. Due to the temporal memoryless property
of the source and sensor observations, we can impose
sample-by-sample estimation without significant estimation
performance loss, but obtain important features such as easy
implementation and minimum delay.

\section{Minimum-Power Estimation with Zero Outage}

In the previous sections we have shown that with Rayleigh fading,
non-zero outage is experienced when there are sum or individual
power constraints. However, with $K$ observation sensors, it is
possible to achieve order-$K$ estimation diversity for both
equal-power and optimal power transmission strategies, while in the
latter case we can further achieve certain adaptive power gain. In
this section, we discuss a converse problem. Given a set of channel
gains, which may be one realization of Rayleigh fading or may be
simply caused by different transmission distances, we seek the
optimal power allocation scheme to minimize the total power
consumption while satisfying a certain distortion requirement. If
the distortion requirement is satisfied with minimum power
consumption for each given channel realization, we call such a
scheme as minimum-power estimation with zero outage.

Based on the above discussion, the minimum-power estimation problem
can be cast as
\begin{eqnarray}
\min && \sum_{k=1}^{K}P_k \nr
\\ \mbox{s.~t.} &&
\Var[\hat{\theta}]\le{D_0}, \nr
\end{eqnarray}
where $D_0$ is the distortion target. According to
Eq.~(\ref{Eq_power1}) and Eq.~(\ref{Eq_distortion}), given a set of
channel SNR $[s_1, s_2, \cdots, s_K]$ and a set of local observation
SNR $[\gamma_1, \gamma_2, \cdots, \gamma_K]$, the above problem is
equivalent to
\begin{eqnarray}
\min_{\alpha_k'} && \sum_{k=1}^{K}\alpha_k'\Wsnr \nr
\\ \mbox{s.~t.} &&
\sgth\left(\sum_{k=1}^{K}\frac{{\alpha_k's_k}}{\gamma_k^{-1}
{\alpha_k's_k}+1}\right)^{-1}\le{D_0} \nonumber
 \\ && \alpha'_k\ge0,
\hspace{0.3cm}k=1,\cdots,K\nr
\end{eqnarray}
Unfortunately, this problem is not convex over the $\alpha_k'$'s.

Let us define
\begin{eqnarray}\label{defr}
r_k=\frac{{\alpha_k's_k}}{\gamma_k^{-1}{\alpha'_ks_k}+1}=\frac{1}
{\gamma_k^{-1}+\frac{1}{\alpha'_ks_k}}, \hspace{0.2cm} \forall k.
\end{eqnarray}
Then the above optimization problem is equivalent to
\begin{eqnarray}
\min_{\alpha_k', r_k} && \sum_{k=1}^{K}\alpha'_k\Wsnr \nr
\\ \mbox{s.~t.} &&
\sum_{k=1}^{K}r_k\ge\frac{\sgth}{D_0} \nr
\\ && r_k=\frac{1}
{\gamma_k^{-1}+\frac{1}{\alpha'_ks_k}},
\hspace{0.3cm}{\alpha'_{k}}\ge0, \hspace{0.3cm}\forall{k}, \nr
\end{eqnarray}
where we see that the variable $\alpha'_k$ can be completely
replaced by a function of $r_k$. From Eq.~(\ref{defr}) we obtain
that
\begin{eqnarray}
\alpha'_k=\frac{1}{s_k(r_k^{-1}-\gamma_k^{-1})}, \hspace{0.2cm}
\forall k.\nr
\end{eqnarray}
Therefore, the problem can be transformed into a problem with
variables $\{r_1, r_2,\dots, r_K\}$ shown as follows (noticing that
$\eta_k:=s_k/\gikpp$):
\begin{eqnarray}\label{S1_final_prob0}
\min_{r_k} &&
\sum_{k=1}^{K}
\frac{1+\gamma_k^{-1}}{s_k(r_k^{-1}-\gamma_k^{-1})}=\sum_{k=1}^{K}
\frac{\eta_k^{-1}}{r_k^{-1}-\gamma_k^{-1}} \nr
\\ \mbox{s.~t.} &&
\sum_{k=1}^Kr_k\ge\frac{\sgth}{D_0};\hspace{0.3cm}
0\le{r_{k}}<\gamma_k,\hspace{0.3cm}\forall{k}
\end{eqnarray}
which is convex over $r_k$. The upper limit on $r_k$ in the second
constraint is due to the fact that $r_k= \frac{1}
{\gamma_k^{-1}+\frac{1}{\alpha'_ks_k}}$ and
$\frac{1}{\alpha'_ks_k}\ge0$.



Similar to solving Eq.~(\ref{Eq_opt}) in
Section~\ref{Section_min_distortion}, the solution of
Eq.~(\ref{S1_final_prob0}) can be stated as follows. As before, we
rank the sensors according to $\eta_1\ge \eta_2\ge\ldots\ge \eta_K$,
and define
\begin{eqnarray}\label{S1_Eq_threshold1}
g(k)=1-\frac{D(k)} {\sqrt{\eta_k}C(k)},\quad \mbox{for\
}1\le{k}\le{K},
\end{eqnarray}
where $C(k)=\displaystyle\sum_{m=1}^{k}
\frac{\gamma_m}{\sqrt{\eta_m}}$ and
$D(k)=\displaystyle\sum_{m=1}^{k}\gamma_m-\frac{\sgth}{D_0}$. Find
$K_1$ such that $g(K_1)>0$ and $g(K_1+1)\le 0$. If $g(k)>0$ for all
$1\le k\le K$, we take $K_1=K$. Also define
$\rho_0={C(K_1)}/{D(K_1)}$. Then the optimal solution is given as
\begin{equation}\label{S1_Eq_result}
r_k^{opt}=\gamma_k \left(1-\sqrt{\eta_k^{-1}}\rho_0\right)^+,
\hspace{0.3cm}\forall\ {k}
\end{equation}
where $(x)^+$ equals $0$ when $x<0$, and otherwise equals $x$.

Hence, by definition, we have
\begin{eqnarray}\label{S1_Eq_powerallocation1}
{\alpha'_k}^{opt}&=&\frac{1}{s_k\left((r_k^{opt})^{-1}-\gamma_k^{-1}\right)}
\nonumber
\\
&=&\left\{\begin{array}{ll}0, & k>K_1\\
\displaystyle\frac{\gamma_k}{s_k}
\left(\sqrt{\eta_k^{-1}}\rho_0-1\right), &k\le
K_1.\end{array}\right.
\end{eqnarray}

Similar to the result in Section~\ref{Section_min_distortion}, we
see that the optimal strategy for minimum-power transmission with
zero outage is to only allocate transmit power to sensors with
better channel SNR and observation quality. Again, the figure of
merit is $\eta_k={s_k}/{(1+\gamma_k^{-1})}$. If a sensor has a
$\eta_k$ below certain threshold, it should be turned off to save
power. Also not surprisingly, the solution in
Eq.~(\ref{S1_Eq_powerallocation1}) is very similar to the one in
Eq.~(\ref{Eq_powerallocation1}) except that the universal constants
$c_0$ and $\rho_0$ are different. In
Eq.~(\ref{Eq_powerallocation1}), $c_0$ is determined by the power
constraint, while in Eq.~(\ref{S1_Eq_powerallocation1}) $\rho_0$ is
determined by the distortion requirement..

\begin{figure}[!h]
\begin{center}
  \scalebox{0.55}{\includegraphics{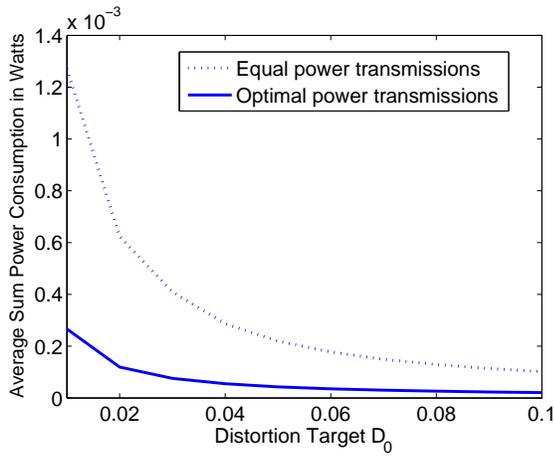}}
\end{center}
 \caption{Average sum power vs. distortion requirements}\label{Fig_power_saving}
\end{figure}

We now solve the optimization problem to show how much power we can
save compared with an equal-power transmission strategy that
satisfies the zero-outage distortion requirement with minimum sum
power. We use the same setup as Section~\ref{Section_equalpower}
except that we now have $100$ sensors and draw the average sum power
consumption over different distortion target values. At each
distortion target $D_0$, the required sum power is averaged over
$10,000$ independent channel realizations. The result is shown in
Fig.~\ref{Fig_power_saving}, where we see that the more strict
distortion requirement we have (smaller $D_0$), the more power we
can save by deploying the optimal power allocation strategies, which
is very important in energy-constrained sensor networks.

{\noindent \bf Discussion of Maximizing Sensor Network Lifetime}

In our model we minimize the power sum $\sum_k{P_k}$, \ie, the
$L^1$-norm of the transmission power vector $\mathbf{P}=[P_1, P_2,
\ldots, P_K]$. If the channel gain and the variance of the
observation noise for each sensor are ergodically time-varying on a
block-by-block basis, minimizing the $L^1$-norm of $\mathbf{P}$ in
each time block minimizes $E\{\sum_k{P_k}\}$. In other words, it
maximizes the lower bound of the average node lifetime that is
defined as $\frac{1}{K}\sum_k\frac{E_0}{E\{P_k\}}$ with $E_0$ the
battery energy available to each sensor (we assume that $E_0$ is the
same for all the sensors). This can be proved from the fact that
$$\frac{1}{K}\sum_k\frac{E_0}{E\{P_k\}}\ge\frac{E_0}{E\{\frac{1}{K}\sum_k{P_k}\}},
$$ which is based on Jensen's inequality~\cite{Boyd1}. However,
when the channel is static and the variance of the observation noise
is time-invariant, minimizing the $L^1$-norm may lead some
individual sensors to consume too much power and die out quickly. In
this case minimizing the $L^\infty$-norm, \ie, minimizing the
maximum of the individual power values, is the fairest for all
sensors, but the total power consumption can be high. A good
compromise is to minimize the $L^2$-norm of
$\mathbf{P}$~\cite{Jinjun_Shuguang1}. In this way, we can penalize
the large terms in the power vector while still keeping the total
power consumption reasonably low. Specifically, for the $L^2$-norm
minimization, the problem formulation becomes
\begin{eqnarray}
\min_{r_k} && \sum_{k=1}^{K}
\left(\frac{\eta_k^{-1}}{r_k^{-1}-\gamma_k^{-1}}\right)^2 \nr
\\ \mbox{s.~t.} &&
\sum_{k=1}^Kr_k\ge\frac{\sgth}{D_0};\hspace{0.3cm}
0\le{r_{k}}<\gamma_k,\hspace{0.3cm}\forall{k}
\end{eqnarray}
which is still a convex problem.
Note that minimizing the various norms of $\mathbf{P}$ may not be
the optimal thing to do given the fact that we are still lack of a
unified definition of sensor network lifetime. A complete
description of this problem is beyond the scope of this paper.
\\

\section{Conclusions}
For the distributed estimation of an unknown source, we have
introduced a new concept of estimation outage and defined the
corresponding estimation diversity, for the case of i.i.d.
observation noise variances at different sensors and i.i.d. fading
channels between the sensors and the fusion center. We have shown
that the full estimation diversity (on the order of the number of
sensor nodes $K$) can be achieved even with simple equal-power
transmission strategies. We have further shown that the end-to-end
distortion can be minimized under sum power constraints, where we
gain certain adaptive power gain on top of the full diversity gain
by turning off sensors with bad channels and bad observation
quality. Moreover, we demonstrated that by considering an extra
individual power constraint at each sensor, certain performance loss
occurs. Minimum-power transmission with zero estimation outage has
also been investigated, where significant power savings is achieved
over equal-power transmission schemes.

\section*{appendix}

\subsection{Derivation of $D_{\infty}$ in Eq.~(\ref{mse_aym_bounds_another})}\label{sec_random_case}

We start with
\begin{eqnarray}
&&\frac{{P_{tot}s_k}}{K\gikpp}-\frac{{P_{tot}s_k}}
{\gamma_k^{-1}{P_{tot}s_k}+K\gikpp} \nr \\
&=&\frac{\gamma_k^{-1}{P_{tot}^2s_k^2}}{K\gikpp
\left(\gamma_k^{-1}{P_{tot}s_k}+K\gikpp\right)} \nr\\
&\le&\frac{\gamma_k^{-1}{P_{tot}^2s_k^2}}{K^2}.\nr
\end{eqnarray}
Thus we have the following inequalities:
\begin{eqnarray}
&&\frac{{P_{tot}s_k}}{K\gikpp}
-\frac{\gamma_k^{-1}{P_{tot}^2s_k^2}}{K^2} \nr \\
&\le&\frac{{P_{tot}s_k}} {\gamma_k^{-1}{P_{tot}s_k}+K\gikpp}
\nr\\
&\le& \frac{{P_{tot}s_k}}{K\gikpp}.\nr
\end{eqnarray}
Therefore, according to Eq.~(\ref{mse_snr}), we have
\begin{eqnarray}\label{mse_appr}
&&\sum_{k=1}^K\frac{{P_{tot}s_k}}{K\gikpp}-
\sum_{k=1}^K\frac{\gamma_k^{-1}{P_{tot}^2s_k^2}}{K^2} \nr\\
 &\le&
\sgth\left(\Var[\hat{\theta}]\right)^{-1} \nr \\
&\le& \sum_{k=1}^K\frac{{P_{tot}s_k}}{K\gikpp}.
\end{eqnarray}

It follows from the strong Law of Large Numbers
(LLN)~\cite{zeitouni} that when $K\to\infty$,
\begin{eqnarray}
\sum_{k=1}^K\frac{{P_{tot}s_k}}{K\gikpp}\to P_{tot}E[s_1/\gionepp], \nr \\
\quad \sum_{k=1}^K\frac{\gamma_k^{-1}{P_{tot}^2s_k^2}}{K}\to
P_{tot}^2 E\left(\gamma_1^{-1}{s_1^2}\right),\nr \\ \quad {\rm
and}\quad \sum_{k=1}^K\frac{\gamma_k^{-1}{P_{tot}^2s_k^2}}{K^2}\to
0, \nr
\end{eqnarray}
providing that $E[s_k/(1+\gamma_k)]$ and $E[\gamma_k^{-1}{s_k^2}]$
are finite.  Therefore,
\begin{eqnarray}\label{mse_aym_bounds}
\lim_{K \rightarrow
\infty}\sgth\left(\Var[\hat{\theta}]\right)^{-1}=
P_{tot}E[s_1/\gionepp],
\end{eqnarray}
which implies that
\begin{eqnarray}
\lim_{K \rightarrow \infty}\Var[\hat{\theta}]
=\frac{\sgth}{P_{tot}E[s_1/\gionepp]}:=D_\infty.
\end{eqnarray}

\subsection{Proof of Theorem~\ref{main_theorem}}

Based on the result from the large deviation theory~\cite{zeitouni},
we first establish the following lemma.

\vspace{1mm}

\begin{lemma}\label{th_ldt}

Suppose $\beta_k: k=1,\ldots, K$ are i.i.d. random variables, and
$a$ is a constant satisfying $a<E(\beta_k)$. Then for any $K\ge 1$,
\begin{eqnarray}\label{exp}
\Prob\left\{\frac{1}{K}\sum_{k=1}^{K}\beta_k
<a\right\}\le\exp(-KI_\beta(a)),
\end{eqnarray}
where $I_\beta(a)$ is the rate function of $\beta_k$ which is
defined to be
\begin{eqnarray}\label{ia}
I_\beta(a)=\sup_{\theta\in\mathbb{R}}(\theta a-\log
M_\beta(\theta)),
\end{eqnarray}
and $M_\beta(\theta)$ is the moment generating function of
$\beta_k$. Similarly, if $a>E(\beta_k)$, then for any $K\ge 1$,
\begin{eqnarray}
\Prob\left\{\frac{1}{K}\sum_{k=1}^{K}\beta_k
>a\right\}\le\exp(-KI_\beta(a)).\nr
\end{eqnarray}

\end{lemma}

\vspace{2mm}

\begin{remark}\label{rmk}
The exponent $I_{\beta}(a)$ in Eq.~(\ref{ia}) is nonnegative (since
$I_{\beta}(a)\ge\left(\theta a-\log
M_\beta(\theta)\right)|_{\theta=0}=0$) and convex over $a$ (since it
is the supremum of linear functions, and is hence convex). Also it
holds that $I_{\beta}(a)=0$ if $a=E(\beta_k)$, $I_{\beta}(a)$ is an
increasing function of $a$ for $a\ge E(\beta_k)$, and is an
decreasing function of $a$ for $a\le E(\beta_k)$. In addition,
$I_{\beta}(a)$ leads to a tight bound in Eq.~(\ref{exp}) in the
sense that
\begin{eqnarray}
\lim_{K\to\infty} -\frac{1}{K}\log
\left(\Prob\left\{\frac{1}{K}\sum_{k=1}^{K}\beta_k
<a\right\}\right)= I_\beta(a)
\end{eqnarray}
if i) $M_\beta(\theta)$ is finite in some neighborhood of $\theta=0$
and ii) $M_\beta(\theta)$ is differentiable in a neighborhood of
$\theta^*$ where the supremum in Eq.~(\ref{ia}) is reached at
$\theta^*$. More details can be found in~\cite{zeitouni} or Section
III of~\cite{weiss}.

\end{remark}

We now continue to prove Theorem~\ref{main_theorem}. From the second
inequality in Eq.~(\ref{mse_appr}), we get
\begin{eqnarray}
{\mathcal{P}_{D_0}}&=&\Prob\left\{\Var[\hat{\theta}]>D_0\right\}\nr\\
&\ge&\Prob\left\{\sum_{k=1}^{K}\frac{P_{tot}s_k}{K\gikpp}
<\frac{\sgth}{D_0}\right\}\nr\\
&=&\Prob\left\{\frac{1}{K}\sum_{k=1}^{K}\frac{s_k}{\gik}
<\frac{\sgth}{D_0P_{tot}}\right\}\nr\\
&=&\Prob\left\{\frac{1}{K}\sum_{k=1}^{K}\frac{s_k}{\gik}
<a\right\},\nr
\end{eqnarray}
where we introduce the constant $a=\sgth/(D_0P_{tot})$. This implies that 
\begin{eqnarray}
&&\lim_{K\to\infty} - \frac{1}{K} \log {\mathcal{P}_{D_0}} \nr \\
&\le& \lim_{K\to\infty}-\frac{1}{K}
\log\left(\Prob\left\{\frac{1}{K}\sum_{k=1}^{K}\frac{s_k}{\gik}
<a\right\}\right).\nr
\end{eqnarray}

On the other hand, from the first inequality in Eq.~(\ref{mse_appr})
we obtain
\begin{eqnarray}
&&\lim_{K\to\infty} -\frac{1}{K} \log {\mathcal{P}_{D_0}} \nr \\
&=&\lim_{K\to\infty} -\frac{1}{K} \log\left(
\Prob\left\{\Var[\hat{\theta}]>D_0\right\}\right)\nr\\
&\ge&\lim_{K\to\infty} -\frac{1}{K}
\log\left(\Prob\left\{\sum_{k=1}^K\frac{{P_{tot}s_k}}{K\gikpp}\right. \right.\nr\\
&&\left.\left.-\sum_{k=1}^K\frac{\gamma_k^{-1}{P_{tot}^2s_k^2}}{K^2}
<\frac{\sgth}{D_0}\right\}\right)\nr\\
&\;{\buildrel {\rm (a)} \over =}\;&\lim_{K\to\infty}-\frac{1}{K}
\log\left(\Prob\left\{\sum_{k=1}^K\frac{{P_{tot}s_k}}{K\gikpp}
<\frac{\sgth}{D_0}\right\}\right)\nr\\
&=&\lim_{K\to\infty}-\frac{1}{K}
\log\left(\Prob\left\{\frac{1}{K}\sum_{k=1}^{K}\frac{s_k}{\gik}<a\right\}\right).\nr
\end{eqnarray}
where $(a)$ is due to the following lemma.

\begin{lemma}\label{lemma_appdx}

Suppose $\{\nu_k: 1\le k\le K\}$, $\{\beta_k: 1\le k\le K\}$ are two
sets of i.i.d. random variables with bounded first moments, $c_1$ is
a constant satisfying $c_1<E(\nu_k)$, and $E(\beta_k)=b$. Further
assume that $\nu_k$ and $\beta_k$ both satisfy the regularity
requirements described in Remark~\ref{rmk}, then
\begin{eqnarray}\label{err_inq}
&&\lim_{K\to\infty}-\frac{1}{K}
\log\left(\Prob\left\{\frac{1}{K}\sum_{k=1}^{K}\nu_k-
\frac{1}{K^2}\sum_{k=1}^{K}\beta_k<c_1\right\}\right)\nr
\\
&=&\lim_{K\to\infty}-\frac{1}{K}\log\left(\Prob\left\{
\frac{1}{K}\sum_{k=1}^{K}\nu_k <c_1\right\}\right).
\end{eqnarray}

\end{lemma}

\vspace{3mm}

\proof We prove $A=B$ by proving $A\ge{B}$ and $A\le{B}$ hold at the
same time. First it is obvious that ``$\le$" holds in
Eq.~(\ref{err_inq}) since
\begin{eqnarray}
&&\Prob\left\{\frac{1}{K}\sum_{k=1}^{K}\nu_k-
\frac{1}{K^2}\sum_{k=1}^{K}\beta_k<c_1\right\} \nr\\  &&\ge
\Prob\left\{ \frac{1}{K}\sum_{k=1}^{K}\nu_k <c_1\right\}.\nr
\end{eqnarray}
We next show the inequality of the other direction. For any
$K\in\mathbb{Z}^+$ and $\epsilon>0$, it holds that
\begin{eqnarray}
&&\Prob\left\{\frac{1}{K}\sum_{k=1}^{K}\nu_k-
\frac{1}{K^2}\sum_{k=1}^{K}\beta_k<c_1\right\}\nr\\
&=&\Prob\left\{\frac{1}{K}\sum_{k=1}^{K}\nu_k-
\frac{1}{K^2}\sum_{k=1}^{K}\beta_k<c_1,\; \frac{1}{K^2}
\sum_{k=1}^{K}\beta_k\le\epsilon\right\} \nr \\
&&+\Prob\left\{\frac{1}{K}\sum_{k=1}^{K}\nu_k-
\frac{1}{K^2}\sum_{k=1}^{K}\beta_k<c_1,\; \frac{1}{K^2}
\sum_{k=1}^{K}\beta_k>\epsilon\right\}
\nr\\
&\le&
\Prob\left\{\frac{1}{K}\sum_{k=1}^{K}\nu_k<c_1+\epsilon\right\}+
\Prob\left\{\frac{1}{K^2}\sum_{k=1}^{K}\beta_k>\epsilon\right\}.\nr
\end{eqnarray}
Taking $\epsilon=b/\sqrt{K}$, we obtain
\begin{eqnarray}\label{eq1}
&&\Prob\left\{\frac{1}{K}\sum_{k=1}^{K}\nu_k-
\frac{1}{K^2}\sum_{k=1}^{K}\beta_k<c_1\right\} \nr \\ &\le&
\Prob\left\{\frac{1}{K}\sum_{k=1}^{K}\nu_k<c_1+\frac{b}{\sqrt{K}}\right\}\nr
\\ && + \Prob\left\{\frac{1}{K}\sum_{k=1}^{K}\beta_k>\sqrt{K}b\right\}.
\end{eqnarray}
Lemma~\ref{th_ldt} implies that
\begin{eqnarray}\label{eq2}
&&\lim_{K\to\infty}-\frac{1}{K}\log\left(\Prob\left\{\frac{1}{K}
\sum_{k=1}^{K}\nu_k<c_1+\frac{b}{\sqrt{K}}\right\}\right)\nr
\\ &=&\lim_{K\to\infty} I_a(c_1+\frac{b}{\sqrt{K}})\nr\\ &=&I_a(c_1),
\end{eqnarray}
where $I_a$ is the rate function of $\nu_k$. Also
\begin{eqnarray}\label{eq3}
&&\lim_{K\to\infty}-\frac{1}{K}\log\left(\Prob\left\{\frac{1}{K}
\sum_{k=1}^{K}\beta_k>\sqrt{K}b\right\}\right) \nr
\\ &=&\lim_{c_2\to\infty}I_b(c_2)\nr \\ &=&+\infty,
\end{eqnarray}
where $I_b$ is the rate function of $\beta_k$.

Therefore, it follows from Eqns~(\ref{eq1})--(\ref{eq3}) that
\begin{eqnarray}
&& \!\!\!\!\!\!\!\!\!\!\!\!\lim_{K\to\infty}-\frac{1}{K}
\log\left(\Prob\left\{\frac{1}{K}\sum_{k=1}^{K}\nu_k-
\frac{1}{K^2}\sum_{k=1}^{K}\beta_k<c_1\right\}\right) \nr \\ &\ge&
I_a(c_1)\nr
\\ &=& \lim_{K\to\infty}-\frac{1}{K} \log\left( \Prob\left\{
\frac{1}{K}\sum_{k=1}^{K}\nu_k <c_1\right\}\right).\nr
\end{eqnarray}
The proof of the lemma is thus completed. \QED

In summary, we have
\begin{eqnarray}
&&\lim_{K\to\infty} - \frac{1}{K} \log {\mathcal{P}_{D_0}} \nr \\
&=& \lim_{K\to\infty}-\frac{1}{K}
\log\left(\Prob\left\{\frac{1}{K}\sum_{k=1}^{K}\frac{s_k}{\gik}
<a\right\}\right).\nr
\end{eqnarray}
Since $s_k/\gikpp$ are i.i.d. random variables, we can apply
Lemma~\ref{th_ldt} to calculate the rate function.


\vspace{2mm}

Applying Lemma \ref{th_ldt} and assuming that the mild regularity
conditions for $M(\theta)$ described in the above remark hold, we
obtain Theorem~\ref{main_theorem}.

\begin{biography}[{\includegraphics[width=1in,height=1.25in,clip,keepaspectratio]
{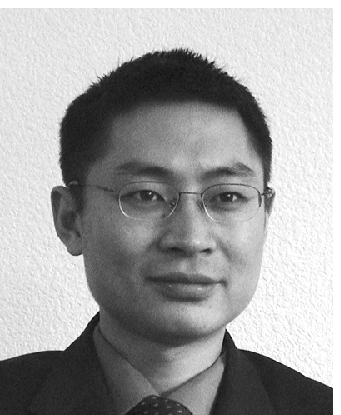}}]{Shuguang Cui} (S'99--M'05) received Ph.D in Electrical
Engineering from Stanford University, California, USA, M.Eng in
Electrical Engineering from McMaster University, Hamilton, Canada,
in 2000, and B.Eng. in Radio Engineering with the highest
distinction from Beijing University of Posts and Telecommunications,
Beijing, China, in 1997. He is now working as an assistant professor
in Electrical and Computer Engineering at the University of Arizona,
Tucson, AZ.

From 1997 to 1998 he worked at Hewlett-Packard, Beijing, P.~R.
China, as a system engineer. In the summer of 2003, he worked at
National Semiconductor, Santa Clara, CA, on the ZigBee project. His
current research interests include cross-layer energy minimization
for low-power sensor networks, hardware and system synergies for
high-performance wireless radios, statistical signal processing, and
general communication theories. He was a recipient of the NSERC
graduate fellowship from the National Science and Engineering
Research Council of Canada and the Canadian Wireless
Telecommunications Association (CWTA) graduate scholarship.

\end{biography}

\vfill

\begin{biography}[{\includegraphics[width=1in,height=1.25in,clip,
keepaspectratio]{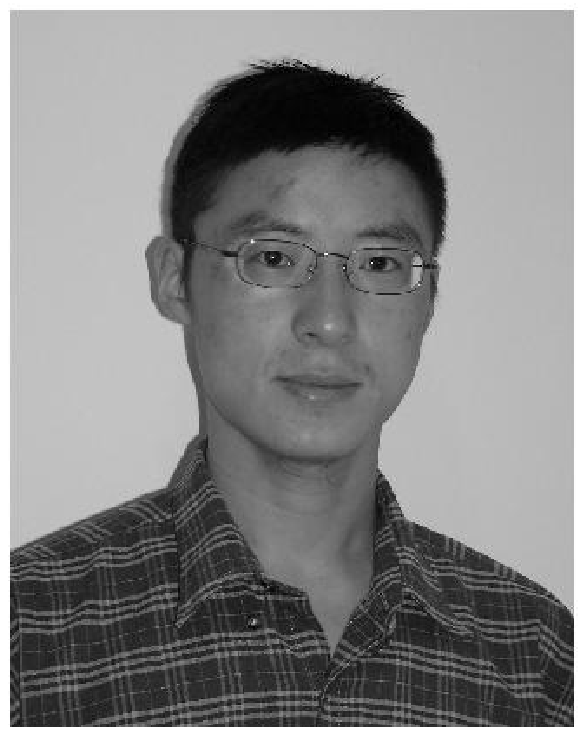}}]{Jin-Jun Xiao}(S'04--M'07) received the
B.Sc.\ degree in Applied Mathematics from Jilin University,
Changchun, China, in 1997, and the M.Sc.\ degree in Mathematics and
Ph.D.\ degree in Electrical Engineering from the University of
Minnesota, Twin Cities, in 2003 and 2006 respectively. He is
currently a Postdoctoral Fellow at the Washington University in St.\
Louis, Missouri.

In the summer of 2002, Dr.\ Xiao worked on barcode signal processing
at Symbol Technologies, Holtsville, New York. His current research
interests are in distributed signal processing, multiuser
information theory and their application to wireless sensor
networks.

\end{biography}

\begin{biography}[{\includegraphics[width=1in,height=1.25in,clip,
keepaspectratio]{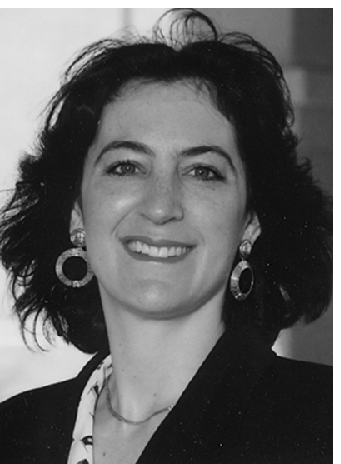}}]{Andrea J. Goldsmith}
(S'90--M'93--SM'99--F'05) is a professor of Electrical Engineering
at Stanford University, and was previously an assistant professor of
Electrical Engineering at Caltech. She has also held industry
positions at Maxim Technologies and at AT$\&$T Bell Laboratories,
and is currently on leave from Stanford as co-founder and CTO of a
wireless startup company. Her research includes work on capacity of
wireless channels and networks, wireless communication theory,
energy-constrained wireless communications, wireless communications
for distributed control, and cross-layer design of wireless
networks. She is author of the book ``Wireless Communications'' and
co-author of the book ``MIMO Wireless Communications,'' both
published by Cambridge University Press. She received the B.S., M.S.
and Ph.D. degrees in Electrical Engineering from U.C. Berkeley.

Dr. Goldsmith is a Fellow of the IEEE and of Stanford, and currently
holds Stanford's Bredt Faculty Development Scholar Chair. She has
received several awards for her research, including the National
Academy of Engineering Gilbreth Lectureship, the Alfred P. Sloan
Fellowship, the Stanford Terman Fellowship, the National Science
Foundation CAREER Development Award, and the Office of Naval
Research Young Investigator Award. She was also a co-recipient of
the 2005 IEEE Communications Society and Information Theory Society
joint paper award. She currently serves as associate editor for the
IEEE Transactions on Information Theory and as editor for the
Journal on Foundations and Trends in Communications and Information
Theory and in Networks. She was previously an editor for the IEEE
Transactions on Communications and for the IEEE Wireless
Communications Magazine, and has served as guest editor for several
IEEE journal and magazine special issues. Dr. Goldsmith is active in
committees and conference organization for the IEEE Information
Theory and Communications Societies, and is currently serving as
technical program co-chair for ISIT 2007. She is an elected member
of the Board of Governers for both societies, a distinguished
lecturer for the IEEE Communications Society, and the second
vice-president and student committee chair of the IEEE Information
Theory Society.

\end{biography}


\begin{biography}[{\includegraphics[width=1in,height=1.25in,clip,
keepaspectratio]{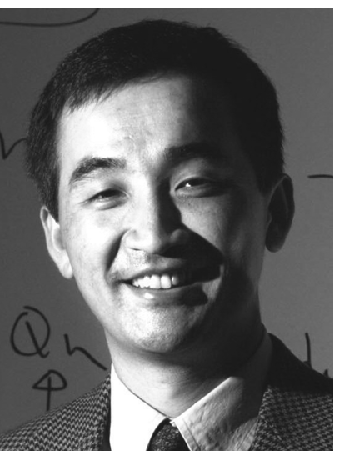}}]{Zhi-Quan (Tom) Luo}(F'07) received his
B.Sc. degree in Applied Mathematics in 1984 from Peking University,
Beijing, China. Subsequently, he was selected by a joint committee
of American Mathematical Society and the Society of Industrial and
Applied Mathematics to pursue Ph.D study in the United States. After
an one-year intensive training in mathematics and English at the
Nankai Institute of Mathematics, Tianjin, China, he entered the
Operations Research Center and the Department of Electrical
Engineering and Computer Science at MIT, where he received a Ph.D
degree in Operations Research in 1989. From 1989 to 2003, Dr. Luo
held a faculty position with the Department of Electrical and
Computer Engineering, McMaster University, Hamilton, Canada, where
he eventually became the department head and held a Canada Research
Chair in Information Processing. Since April of 2003, he has been
with the Department of Electrical and Computer Engineering at the
University of Minnesota (Twin Cities) as a full professor and holds
an endowed ADC Chair in digital technology. His research interests
lie in the union of optimization algorithms, data communication and
signal processing.

Prof. Luo serves on the IEEE Signal Processing Society Technical
Committees on Signal Processing Theory and Methods (SPTM), and on
the Signal Processing for Communications (SPCOM). He is a co-reciept
of the 2004 IEEE Signal Processing Society's Best Paper Award, and
has held editorial positions for several international journals
including Journal of Optimization Theory and Applications,
Mathematics of Computation, and IEEE Transactions on Signal
Processing. He currently serves on the editorial boards for a number
of international journals including SIAM Journal on Optimization,
Mathemtical Programming, and  Mathematics of Operations Research.

\end{biography}

\begin{biography}[{\includegraphics[width=1in,height=1.25in,clip,
keepaspectratio]{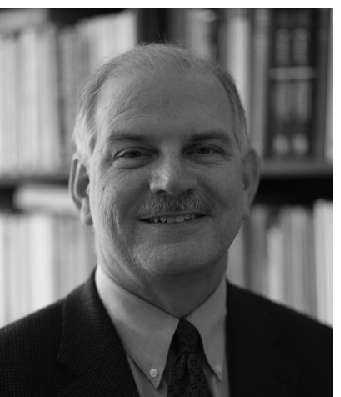}}]{H. Vincent Poor}
(S'72--M'77--SM'82--F'87) received the Ph.D. degree in EECS from
Princeton University in 1977.  From 1977 until 1990, he was on the
faculty of the University of Illinois at Urbana-Champaign. Since
1990 he has been on the faculty at Princeton, where he is the Dean
of Engineering and Applied Science, and the Michael Henry Strater
University Professor of Electrical Engineering. Dr. Poor's research
interests are in the areas of stochastic analysis, statistical
signal processing, and their applications in wireless networks and
related fields. Among his publications in these areas is the recent
book MIMO Wireless Communications (Cambridge University Press,
2007).

Dr. Poor is a member of the National Academy of Engineering and is a
Fellow of the American Academy of Arts and Sciences. He is also a
Fellow of the Institute of Mathematical Statistics, the Optical
Society of America, and other organizations.  In 1990, he served as
President of the IEEE Information Theory Society, and he is
currently serving as the Editor-in-Chief of the IEEE Transaction on
Information Theory. Recent recognition of his work includes a
Guggenheim Fellowship and the IEEE Education Medal.

\end{biography}

\end{document}